\documentclass[prc,showpacs,nofootinbib,preprint]{revtex4-1}
\usepackage{epsfig,graphics,color}
\usepackage{subfigure}
\usepackage{bbm}
\usepackage{amssymb}
\usepackage{amsmath}

\newcommand{\bra}[1]{\langle #1|}
\newcommand{\ket}[1]{|#1\rangle}
\newcommand{\braket}[2]{\langle #1|#2\rangle}
\newcommand{\roundbra}[1]{(#1|}
\newcommand{\roundket}[1]{|#1)}

\def\one{{\bf 1}\,}
\def\quarknumberoperator{{\mathbbm 1}\,}

\def\Dslash{D \hspace{-2.7mm}/ \;}

\def\w2{\tilde w^2}
\def\ws2{1}

\newcommand{\chiral}[1]{\stackrel{\circ}{#1}}

\begin{document}
\title{Large-$N_c$ operator analysis \\of 2-body meson-baryon counterterms in the chiral Lagrangian }
\author{M.F.M. Lutz and A. Semke}
\affiliation{Gesellschaft f\"ur Schwerionenforschung (GSI),\\
Planck Str. 1, 64291 Darmstadt, Germany}
\date{\today}
\begin{abstract}
The chiral $SU(3)$ Lagrangian with the baryon octet and decuplet fields is considered. The $Q^2$ counterterms involving
the decuplet fields are constructed. We derive the correlation of the chiral parameters implied by the $1/N_c$ expansion at
leading order in QCD.
\end{abstract}

\maketitle

\section{Introduction}

The chiral $SU(3)$ Lagrangian with the baryon octet and decuplet fields has been used in the
literature in application to the chiral expansion of the
baryon masses \cite{Bernard:Kaiser:Meissner:1993,Borasoy:Meissner:1997,Donoghue:Holstein:Borasoy:1999,Borasoy:1999,Borasoy-et-al:2002,Pascalutsa:Vanderhaeghen:2005,Semke:Lutz:2005,Semke:Lutz:2006}.
So far the sector of the chiral Lagrangian involving the decuplet fields is much less explored as compared to the sector involving the
baryon octet fields.

A new type of application was suggested in Ref.~\cite{Copenhagen} where the leading order term involving the decuplet fields was used as the
driving term in a coupled-channel computation. It was shown that the s-wave interaction of the Goldstone bosons with the decuplet states
leads to the formation of a resonance spectrum with $J^P=\frac{3}{2}^-$ quantum numbers reasonably close to the empirical one. In order to
improve such an approach it is important to consider chiral correction terms. The purpose of the present work is
to study the $Q^2$ counterterms involving the baryon decuplet fields systematically. In order to
reduce the number of unknown parameters, the $1/N_c$ expansion is performed \cite{large-N_c-reference}. We shall apply the
technology developed in Refs.~\cite{Luty:Russel:1994,Dashen:Jenkins:Manohar:1994}. A rigorous correlation of the $Q^2$ parameters is achieved
upon the evaluation of baryon matrix elements of current-current correlation functions. The latter are expanded in powers of $1/N_c$
applying the operator reduction rules of Ref.~\cite{Dashen:Jenkins:Manohar:1994}.

All together we find  25 symmetry preserving terms relevant at order $Q^2$. At leading order in the $1/N_c$ expansion we derive 18
sum rules, which reduce the number of free parameters to 7.

\section{Chiral Lagrangian with baryon octet and decuplet fields} \label{section:chiral-lagrangian}

The construction rules for the chiral $SU(3)$ Lagrangian density are recalled.
For more technical details see for example Refs.~\cite{Weinberg68,GL84,Krause:1990,Kaym84,Ecker89,Borasoy,Birse,Becher,Fuchs}. 
The basic building blocks of the chiral Lagrangian  are
\begin{eqnarray}
&& U_\mu = {\textstyle{1\over 2}}\,e^{-i\,\frac{\Phi}{2\,f}} \left(
\partial_\mu \,e^{i\,\frac{\Phi}{f}} \right) e^{-i\,\frac{\Phi}{2\,f}}
-{\textstyle{i\over 2}}\,e^{-i\,\frac{\Phi}{2\,f}} \,r_\mu\, e^{+i\,\frac{\Phi}{2\,f}}
+{\textstyle{i\over 2}}\,e^{+i\,\frac{\Phi}{2\,f}} \,l_\mu\, e^{-i\,\frac{\Phi}{2\,f}} \,,
\nonumber\\
&& f_{\mu \nu}^{\pm} = \frac{1}{2}\,e^{+i\,\frac{\Phi}{2\,f}} \left( \partial_\mu \,l_\nu- \partial_\nu \,l_\nu -i\,[l_\mu ,\,l_\nu] \right)
e^{-i\,\frac{\Phi}{2\,f}}
\nonumber\\
&&  \quad \;\;\,\pm\, \frac{1}{2}\,e^{-i\,\frac{\Phi}{2\,f}} \left( \partial_\mu \,r_\nu\,- \partial_\nu \,r_\nu \,-i\,[r_\mu \,,\,r_\nu\,] \right)
e^{+i\,\frac{\Phi}{2\,f}} \,,\quad  B \;, \quad B_\mu  \;,
\label{def-fields}
\end{eqnarray}
where we include the pseudo-scalar meson octet fields
$\Phi(J^P\!\!=\!0^-)$,
the baryon octet fields $B(J^P\!\!=\!{\textstyle{1\over2}}^+)$ and the baryon
decuplet fields $B_\mu(J^P\!\!=\!{\textstyle{3\over2}}^+)$.  The classical source functions $r_\mu$ and $l_\mu$ in Eq.~(\ref{def-fields}) are
linear combinations of the vector and axial-vector sources with $r_\mu = v_\mu+a_\mu$ and $l_\mu = v_\mu-a_\mu$.

The octet fields may be decomposed into their isospin multiplets,
\begin{eqnarray}
&& \Phi = \tau \cdot \pi
+ \alpha^\dagger \cdot  K +  K^\dagger \cdot \alpha
+ \eta\,\lambda_8\,,
\nonumber\\
&& \sqrt{2}\,B =  \alpha^\dagger \!\cdot \! N + \lambda_8 \,\Lambda + \tau \cdot \Sigma
 +\Xi^t\,i\,\sigma_2 \!\cdot \!\alpha   \, ,
\nonumber\\
&&\; \alpha^\dagger = {\textstyle{1\over \sqrt{2}}}\,(\lambda_4+i\,\lambda_5 ,\lambda_6+i\,\lambda_7 )\,,\qquad
\tau = (\lambda_1,\lambda_2, \lambda_3)\,,
\label{def-octet-fields}
\end{eqnarray}
with $K=(K^{(+)},K^{(0)})^t$,  for instance. The matrices $\lambda_i$ are the Gell-Mann matrices and $\sigma_2$ is the second Pauli matrix.
The baryon decuplet fields $B^{abc}_\mu $ are completely symmetric in $a,b,c=1,2,3$ and are related to the physical states by
\begin{eqnarray}
\begin{array}{llll}
B^{111} = \Delta^{++}\,, & B^{113} =\Sigma^{+}/\sqrt{3}\,, &
B^{133}=\Xi^0/\sqrt{3}\,,  &B^{333}= \Omega^-\,, \\
B^{112} =\Delta^{+}/\sqrt{3}\,, & B^{123} =\Sigma^{0}/\sqrt{6}\,, &
B^{233}=\Xi^-/\sqrt{3}\,, & \\
B^{122} =\Delta^{0}/\sqrt{3}\,, & B^{223} =\Sigma^{-}/\sqrt{3}\,, &
& \\
B^{222} =\Delta^{-}\,. & & &
\end{array}
\label{def-decuplet-fields}
\end{eqnarray}
The parameter $f $ in Eq.~(\ref{def-fields}) may be obtained from the weak decay widths of the charged pions and kaons with
$f_\pi = (92.42 \pm 0.33) $ MeV and $f_K =( 113.0 \pm 1.3 )$ MeV \cite{fpi:exp}. The detailed analysis
in Ref.~\cite{GL85} derives $f_\pi/f = 1.07 \pm 0.12$ from an extrapolation of $f_\pi $ and $f_K$ to their value in the chiral limit $f$.

Explicit chiral symmetry-breaking is included in terms
of scalar and pseudo-scalar source fields $\chi_\pm $ proportional to the quark mass
matrix of QCD
\begin{eqnarray}
\chi_\pm = \frac{1}{2} \left(
e^{+i\,\frac{\Phi}{2\,f}} \,\chi_0 \,e^{+i\,\frac{\Phi}{2\,f}}
\pm e^{-i\,\frac{\Phi}{2\,f}} \,\chi_0 \,e^{-i\,\frac{\Phi}{2\,f}}
\right) \,,
\label{def-chi}
\end{eqnarray}
with $\chi_0 =2\,B_0\, {\rm diag} (m_u,m_d,m_s)$.
The merit of the particular field combinations in Eqs.~(\ref{def-fields}) and (\ref{def-chi}) is their identical transformation property under the chiral $SU_L(3)\otimes SU_R(3)$ rotations.

The chiral Lagrangian consists of all possible interaction
terms, formed with the fields $U_\mu,\, B,\, B_\mu, \, f_{\mu\nu}^\pm $ and $\chi_\pm$.
Derivatives of the fields must be included in compliance with the local chiral $SU(3)$ symmetry. This leads to the notion of a covariant derivative $D_\mu$ which is identical for all fields in Eqs.~(\ref{def-fields}) and (\ref{def-chi}). For baryons, the covariant derivative $D_\mu$  acts on the octet and decuplet fields as follows:
\begin{eqnarray}
&&(D_\mu  B)^{a}_{b} = \partial_\mu B^{a}_{b} +  \Gamma^{a}_{\mu,l}\, B^{l}_{b} -
\Gamma^{l}_{\mu,b}\, B^{a}_{l} \,, \quad
\nonumber\\
&& (D_\mu B_\nu)^{abc} = \partial_\mu B^{abc}_\nu + \Gamma^{a}_{\mu,l } B^{lbc}_{\nu}
+ \Gamma^{b}_{\mu,l } B^{alc}_{\nu}+\Gamma^{c}_{\mu,l } B^{abl}_{\nu} \,,
\label{def-covariant-derivative}
\end{eqnarray}
with $\Gamma_\mu=-\Gamma_\mu^\dagger$ given by
\begin{eqnarray*}
 &&\Gamma_\mu ={\textstyle{1\over 2}}\,e^{-i\,\frac{\Phi}{2\,f}} \,
\Big[\partial_\mu -i\,(v_\mu + a_\mu) \Big] \,e^{+i\,\frac{\Phi}{2\,f}}
+{\textstyle{1\over 2}}\, e^{+i\,\frac{\Phi}{2\,f}} \,
\Big[\partial_\mu -i\,(v_\mu - a_\mu)\Big] \,e^{-i\,\frac{\Phi}{2\,f}}\,.
\end{eqnarray*}

To cope with flavor indices in the products of SU(3) tensors containing decuplet fields, we shall apply a compact \textit{dot-notation}
suggested in Ref.~\cite{Lutz:Kolomeitsev:2002}:
\begin{eqnarray}\label{def:dot-notation}
(\bar B^\mu \cdot B_\nu)^m_k \equiv \bar B^\mu_{ijk} \,B_\nu^{ijm}, \quad (\bar B^\mu \cdot \Phi)^m_k \equiv \bar B^\mu_{ijk}\, \Phi^i_l\, \epsilon^{jlm}, \quad (\Phi \cdot B_\mu)^m_k \equiv B_\mu^{ijm}\, \Phi_i^l\, \epsilon_{jlk}\,.
\end{eqnarray}

The chiral Lagrangian is a powerful tool, once it is combined with appropriate
counting rules leading to a systematic approximation strategy.
We aim at describing hadronic interactions at low-energy by constructing an expansion
in small momenta and small quark masses and by supplementing this interaction with a parametric $1/N_c$ expansion.

The leading order chiral Lagrangian is of the order $Q$ and is given by
\begin{eqnarray}\label{lbl:chiral_L_Q1}
\mathcal{L}^{(1)} &=&
\mathrm{tr}\, \Big\{ \bar B (i\, \Dslash\, - \chiral M_{[8]})\, B \Big\}
+ F\, \mathrm{tr} \Big\{ \bar{B}\, \gamma^\mu \gamma_5\, [i\,U_\mu,B]\, \Big\} + D\, \mathrm{tr}\Big\{ \bar{B}\, \gamma^\mu \gamma_5\, \{i\,U_\mu,\,B\}\, \Big\}
\nonumber \\
&-& \mathrm{tr}\, \Big\{ \bar B_\mu \cdot \big((i\,\Dslash\, - \chiral M_{[10]})\,g^{\mu\nu} -i\,(\gamma^\mu D^\nu + \gamma^\nu D^\mu) + \gamma^\mu(i\,\Dslash + \chiral M_{[10]})\gamma^\nu \big)\, B_\nu \Big\}
\nonumber \\
&+& C\left( \mathrm{tr} \Big\{ (\bar{B}_\mu \cdot i\, U^\mu)\, B\Big\} + \mathrm{h.c.} \right) + H\, \mathrm{tr} \Big\{ (\bar{B}^\mu\cdot \gamma_\nu  \gamma_5\, B_\mu)\, i\,U^\nu \Big\}.
\end{eqnarray}
The baryon masses in the chiral limit are given by the values of $\chiral M_{[8]}$ and $\chiral M_{[10]}$ for the members of the flavor $SU(3)$ octet and decuplet, respectively.
The parameters $F$ and $D$ in Eq.~(\ref{lbl:chiral_L_Q1}) may be determined from the study of semi-leptonic decays of baryons, $B\rightarrow B^\prime + e + \bar \nu_e$. The value of $C$ may be extracted from the hadronic decays of the decuplet baryons. Loop corrections to these decays allow an estimate of the parameter $H$ \cite{Butler1993}. Using large-$N_c$ sum rules, the parameters $C$ and $H$ may be also estimated given the empirical values for
$F$ and $D$ \cite{Dashen}.

Our main purpose is the study of the chiral symmetry conserving $Q^2$ counterterms. These terms were constructed systematically for the baryon octet fields (see e.g. \cite{CH-Lee,Lutz:Kolomeitsev:2002}). The analogous terms involving the decuplet fields were not presented in literature. Some partial results can be found in Ref.~\cite{Tiburzi2004}.

It is convenient to group the $Q^2$ counterterms according to their Dirac structure into
scalar, vector, axial-vector and tensor terms:
\begin{eqnarray}\label{lbl:L_MB_four_point}
\mathcal{L}^{(2)}=\mathcal{L}^{(S)} + \mathcal{L}^{(V)} + \mathcal{L}^{(A)}+ \mathcal{L}^{(T)}\,,
\end{eqnarray}
with
\begin{eqnarray}
\mathcal{L}^{(S)} &=&- \frac{1}{2}\,g_0^{(S)}\,\mathrm{tr} \,\Big\{\bar{B}\,B \Big\}\, \mathrm{tr}\Big\{ U_\mu\, U^\mu \Big\} - \frac{1}{2}\,g_1^{(S)}\,\mathrm{tr} \,\Big\{ \bar{B}\, U^\mu \Big\}\, \mathrm{tr}\,\Big\{U_\mu\, B \Big\}
\nonumber \\
&-&\frac{1}{4}\,g_D^{(S)} \mathrm{tr}\,\Big\{\bar{B}\left\{\left\{U_\mu, U^\mu\right\}, B\right\}\Big\}
-\frac{1}{4}\,g_F^{(S)}\mathrm{tr}\,\Big\{ \bar{B}\left[\left\{U_\mu, U^\mu\right\}, B\right]\Big\}
\nonumber\\
&+& \frac{1}{2}\,h_1^{(S)}\,\mathrm{tr}\,\Big\{ \bar{B}_\mu \cdot B^\mu \Big\}\, \mathrm{tr}\,\Big\{U_\nu\; U^\nu\Big\} +
\frac{1}{2}\,h_2^{(S)}\,\mathrm{tr}\,\Big\{\bar{B}_\mu \cdot B^\nu \Big\}\, \mathrm{tr}\,\Big\{U^\mu\, U_\nu\Big\}
\nonumber \\
&+& h_3^{(S)}\,\mathrm{tr}\,\Big\{\Big(\bar{B}_\mu \cdot B^\mu\Big)\, \Big(U^\nu\, U_\nu\Big) \Big\} + \frac{1}{2}\,h_4^{(S)}\,\mathrm{tr}\,\Big\{ \Big(\bar{B}_\mu \cdot B^\nu\Big)\, \{U^\mu,\, U_\nu \} \Big\}
\nonumber \\
&+& h_5^{(S)}\, \mathrm{tr}\, \Big\{ \Big( \bar{B}_\mu \cdot U_\nu\Big)\, \Big(U^\nu\cdot B^\mu \Big) \Big\}
\nonumber \\
&+& \frac{1}{2}\,h_6^{(S)}\, \mathrm{tr} \Big\{ \Big( \bar{B}_\mu \cdot U^\mu\Big)\, \Big(U^\nu\cdot B_\nu \Big)
+\Big( \bar{B}_\mu \cdot U^\nu\Big)\, \Big(U^\mu\cdot B_\nu \Big) \Big\} \, ,
\nonumber\\
\mathcal{L}^{(V)} &=& -\frac{1}{4}\,g_0^{(V)}\, \Big( \mathrm{tr}\,\Big\{\bar{B}\, i\,\gamma^\mu\, \partial^\nu B\Big\} \,
\mathrm{tr}\,\Big\{ U_\nu\, U_\mu \Big\} + \mathrm{h.c.} \Big)
\nonumber \\
&-& \frac{1}{8}\,g_1^{(V)} \,\Big( \mathrm{tr}\,\Big\{\bar{B}\,U_\mu \Big\} \,i\,\gamma^\mu \, \mathrm{tr}\,\Big\{U_\nu\, \partial^\nu B\Big\} + \mathrm{tr}\,\Big\{\bar{B}\,U_\nu \Big\} \,i\,\gamma^\mu \, \mathrm{tr}\,\Big\{U_\mu\, \partial^\nu B\Big\} + \mathrm{h.c.} \Big)
\nonumber \\
&-& \frac{1}{8}\,g_D^{(V)}\, \Big(\mathrm{tr}\,\Big\{\bar{B}\, i\,\gamma^\mu \left\{\left\{U_\mu,\, U_\nu\right\}, \partial^\nu B\right\}\Big\} + \mathrm{h.c.} \Big)
\nonumber\\
&-& \frac{1}{8}\,g_F^{(V)}\,\Big( \mathrm{tr}\,\Big\{ \bar{B}\, i\,\gamma^\mu\, \left[\left\{U_\mu,\, U_\nu\right\},\, \partial^\nu B \right]\Big\} + \mathrm{h.c.} \Big)
\nonumber \\
&+& \frac{1}{4}\,h_1^{(V)}\,\Big(\mathrm{tr}\,\Big\{ \bar{B}_\lambda \cdot i\,\gamma^\mu\, \partial^\nu B^\lambda\Big\} \,\mathrm{tr}\,\Big\{U_\mu\, U_\nu\Big\} + \mathrm{h.c.}\Big)
\nonumber \\
&+& \frac{1}{4}\,h_2^{(V)}\,\Big(\mathrm{tr}\,\Big\{ \left(\bar{B}_\lambda \cdot i\,\gamma^\mu\, \partial^\nu B^\lambda \right) \{U_\mu,\, U_\nu\}\Big\} + \mathrm{h.c.}\Big)
\nonumber \\
&+& \frac{1}{4}\,h_3^{(V)}\, \Big( \mathrm{tr}\, \Big\{ \left( \bar{B}_\lambda \cdot U_\mu\right) i\,\gamma^\mu \left(U_\nu\cdot \partial^\nu B^\lambda \right) + \left( \bar{B}_\lambda \cdot U_\nu\right) i\,\gamma^\mu \left(U_\mu\cdot \partial^\nu B^\lambda \right) \Big\} + \mathrm{h.c.}\Big), \nonumber \\
\mathcal{L}^{(A)} &=& -\frac{1}{4}\,f_1^{(A)}\, \Big( \mathrm{tr} \,\Big\{ (\bar B^\mu \cdot \gamma^\nu \gamma_5\,  B)\, \{U_\mu,\, U_\nu\} \Big\} + \mathrm{h.c.}\Big)
\nonumber \\
&-& \frac{1}{4}\,f_2^{(A)}\, \Big( \mathrm{tr} \,\Big\{ (\bar B^\mu \cdot \gamma^\nu\,\gamma_5\, B)\, [U_\mu,\, U_\nu] \Big\}
+ \mathrm{h.c.}\Big)  \nonumber \\
&-& \frac{1}{4}\,f_3^{(A)}\, \Big(\mathrm{tr} \,\Big\{ (\bar B^\mu \cdot U_\nu)\, \gamma^\nu \gamma_5\, (U_\mu \cdot B)
+(\bar B^\mu \cdot U_\mu)\,\gamma^\nu \,\gamma_5\, (U_\nu \cdot B) \Big\} + \mathrm{h.c.} \Big)
\nonumber \\
&-& \frac{1}{4}\,f_4^{(A)}\, \Big(\mathrm{tr} \,\Big\{ (\bar B^\mu \cdot U_\nu)\, \gamma^\nu \gamma_5\, (U_\mu \cdot B)  -
 (\bar B^\mu \cdot U_\mu)\, \gamma^\nu \,\gamma_5\, (U_\nu \cdot B) \Big\} + \mathrm{h.c.} \Big) \,,
\nonumber\\
\mathcal{L}^{(T)} &=& -\frac{1}{2}\,g_1^{(T)}\, \mathrm{tr}\,\Big\{ \bar{B}\, U_\mu\Big\}\, i\,\sigma^{\mu\nu}\,
 \mathrm{tr}\,\Big\{U_\nu\, B\Big\}
- \frac{1}{4}\,g_D^{(T)} \mathrm{tr}\,\Big\{ \bar{B}\, i\,\sigma^{\mu\nu}\, \left\{\left[U_\mu, U_\nu\right],\, B\right\} \Big\}
\nonumber \\
&-& \frac{1}{4}\,g_F^{(T)}\, \mathrm{tr}\,\Big\{ \bar{B}\, i\,\sigma^{\mu\nu}\, \left[\left[U_\mu, U_\nu\right],\, B\right] \Big\}
 \nonumber \\
&+& \frac{1}{2}\,h_1^{(T)}\,\mathrm{tr}\,\Big\{ \left(\bar{B}_\alpha \cdot i\,\sigma^{\mu \nu}\,B^\alpha\right)
\left[U_\mu ,\, U_\nu \right] \Big\}  \, .
\label{def-Q2-terms}
\end{eqnarray}
Though the spin-3/2 fields introduce additional Lorentz structures, the on-shell conditions $\gamma_\mu\,B^\mu =0$ and
$\partial_\mu \,B^\mu=0$ reduce the number of relevant terms significantly.
We affirm that a complete list of $Q^2$ terms is presented in Eq.~(\ref{def-Q2-terms}).

\section{Large-$N_c$ operator analysis }

In deriving the large-$N_c$ sum rules that correlate the parameters of the $Q^2$ counterterms as introduced in the previous section, we
follow the work of Luty and March-Russell \cite{Luty:Russel:1994}.  Their approach is based on the description of the ground-state baryons
in terms of localized mean-field quarks. Quarks are put into a coordinate-dependent scalar potential wall. The average field seen by every
quark is constructed in such a way as to localize the lowest quark state at a typical scale $\Lambda_{QCD}$. This can be achieved by a position-dependent
quark mass matrix $m_q(|\vec x|)$.  The full QCD Hamilton operator, $H$, may be split into a
mean-field part, $H_0$, and an interaction part, $V$, without changing the structure of $H$:
\begin{eqnarray}
H = H_0 + V\,.
\label{def:full-H_with-mean-field}
\end{eqnarray}
This procedure, which should be regarded just as a mathematical trick, reorganizes the
perturbative expansion and helps to study the structure of matrix elements in the large-$N_c$ limit.
The ``free'' quark-field operator, $\Psi_I(t, \vec x)$, can be expanded in the eigenmodes of the $H_0$
as follows:
\begin{eqnarray}
\Psi_I(t, \vec x) = \sum_{\alpha =1}^2 \sum_{n=0}^\infty \Big( u_{n,\, \alpha}(\vec x)\,e^{-\,i\,E_n\, t}\, b_{n,\, \alpha}
+ v_{n,\,\alpha}(\vec x)\, e^{+\,i\,E_n\, t}\, d^\dagger_{n,\,\alpha}\Big)\,.
\label{def:quark-field-operator-mean-field}
\end{eqnarray}
Here, $b_{n,\, a}$ destroys a quark in the $n$th mode with the spin quantum numbers $\alpha$ and $d^\dagger_{n,\, \alpha}$ creates
an anti-quark with the quantum numbers $n$ and $\alpha $,  $E_n$ is the corresponding energy eigenvalue.
The object of main interest are the \textit{baryonic ground states} which are the lowest color-neutral eigenstates of the Hamilton operator
$H_0$ in Eq.~(\ref{def:full-H_with-mean-field}) with baryon number one,
\begin{eqnarray}
\ket{\mathcal B_0}= \mathcal B^{\alpha_1 a_1\ldots \alpha_{N_c} a_{N_c}}\,
\varepsilon^{A_1\ldots A_{N_c}}\, b^\dagger_{0,\, \alpha_1 a_1 A_1}\cdots b^\dagger_{0,\, \alpha_{N_c} a_{N_c} A_{N_c}} \ket 0\,,
\label{def:baryon-ground-state}
\end{eqnarray}
with the spin indices $\alpha_i=1,2$, the flavor indices $a_i=1,2,3$ and the color indices $A_i=1, \ldots N_c$.
The color structure of these states is completely furnished by the antisymmetric tensor $\varepsilon$, whereas the ``wave function''
$\mathcal B$ specifies their spin and flavor quantum numbers.
The state $\ket{\mathcal B_0}$ is a tensor product of $N_c$ one-quark states, created by the quark creation
operators $b^\dagger_{0,\, \alpha a A}$ acting on the ground state, $\ket 0$, of $H_0$. It is always possible to adjust the scalar mean
field discussed above in such a way as to make the lightest mode component describing a quark state localized at the scale
$\Lambda_{QCD}$.

The quark field operator in the Heisenberg picture follows by a unitary transformation from
its interaction picture representation in Eq.~(\ref{def:quark-field-operator-mean-field}) as follows
\begin{eqnarray}
\Psi(t,\vec x)=U^\dagger (t)\, \Psi_I(t,\vec x)\, U(t), \qquad U(t)= e^{+i\,H_0\,t}\,e^{-i\,H\,t}=\mathcal T \exp\left[-i\int^{t}_0 dt^\prime\, V_I(t^\prime) \right]\,,
\label{def:Heisenberg-field}
\end{eqnarray}
with $V_I(t)= e^{+i\,H_0\,t}\,V\,e^{-i\,H_0\,t}$. The states $\ket{\mathcal B_0}$ evolve to the eigenstates of the full
Hamiltonian, $\ket{\mathcal B}$, with the energy $E_\mathcal B$ according to
\begin{eqnarray}
e^{-i\,H\,T} \,\ket{\mathcal B_0} = \ket{\mathcal B}\braket{\mathcal B}{\mathcal B_0} e^{-i\,E_{\mathcal B} \,T} + \ldots \,,
 \label{def:baryon-state-time-evolution}
\end{eqnarray}
where the states $\ket{\mathcal B}$ and $\ket{\mathcal B_0}$ have identical quantum numbers. The terms indicated by
ellipses in Eq.~(\ref{def:baryon-state-time-evolution}) correspond to the states with energies $E>E_{\mathcal B}$ and are suppressed for $T \rightarrow \infty (1-i\,\epsilon)$.

We consider baryon matrix elements of time-ordered products of Heisenberg operators, $\mathcal{O}_{i}(x_i)$.
With the help of Eq.~(\ref{def:baryon-state-time-evolution}) such matrix elements may be expressed in terms of the
operator in the interaction picture, $\mathcal{O}_{I i}(x_i)$, and the bare baryon states introduced in Eq.~(\ref{def:baryon-ground-state}). It holds
\begin{eqnarray}
&&\bra{\mathcal{B}^\prime} \mathcal T\, \mathcal{O}_{1}(x_1) \ldots  \mathcal{O}_{n}(x_n) \ket{\mathcal{B}} =
\frac{ \bra{\mathcal{B}^\prime_0}\,\mathcal T\, \mathcal{O}_{I\,1}(x_1) \ldots \mathcal{O}_{I\,n}(x_n)\, U_I \,
\ket{\mathcal{B}_0} } {\bra{\mathcal{B}^\prime_0}\, U_I\, \ket{\mathcal{B}^\prime_0}^{1/2}\, \bra{\mathcal{B}_0}\,
U_I\, \ket{\mathcal{B}_0}^{1/2} } +\cdots \,,
\nonumber\\
&& U(-T)\,\ket{\mathcal{B}} = \frac{\ket{\mathcal{B}_0}}{\bra{\mathcal{B}_0}\,
U_I\, \ket{\mathcal{B}_0}^{1/2}} + \cdots \,, \qquad \quad
 U_I = \mathcal T\, \exp \left[-i\int^{+T}_{-T} dt^\prime\, V_I(t^\prime) \right] \,,
 \label{def:QCD-Green-function}
\end{eqnarray}
where the time $T$ must be larger than the modulus of all time arguments $|t_i| \ll T$
with $i=1, \cdots, n$.  For sufficiently large $T$ the additional terms indicated by ellipses in Eq.~(\ref{def:QCD-Green-function}) turn
irrelevant \cite{Luty:Russel:1994}.

\begin{figure}[b]
\centering
\subfigure[Expansion in terms of quark-gluon vertices]{\includegraphics[width=12cm,clip=true]{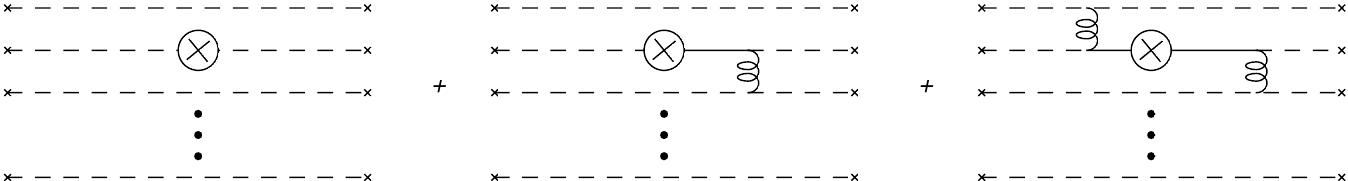} \label{fig:QCD_operator_me_expansion_a} }
\subfigure[Expansion in terms of effective vertices]{\includegraphics[width=12cm,clip=true]{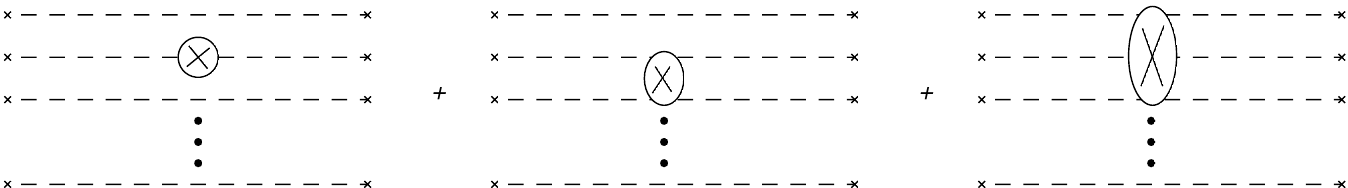}\label{fig:QCD_operator_me_expansion_b}}
\caption{Diagrammatic representation of the matrix elements of an one-body operator in Eq.~(\ref{def:QCD-Green-function}).}
\label{fig:QCD_operator_me_expansion}
\end{figure}

The r.h.s. of Eq.~(\ref{def:QCD-Green-function}) can be analyzed in perturbation theory by appropriately defined Feynman rules. The time ordered
product is written as a sum of all contractions and products of normal ordered operators. The denominator
in Eq.~(\ref{def:QCD-Green-function}) cancels those terms in the expansion, where all the operators are contracted \cite{Luty:Russel:1994}.
Normal ordered products do contribute, because the matrix elements are taken between non-vacuum states. Since the baryon state
$\ket{\mathcal{B}_0}$ is generated by $N_c$ quark creation operators $b_0^\dagger$, the matrix element in Eq.~(\ref{def:QCD-Green-function})
singles out only those normal ordered contributions that are characterized by $N_c$ annihilation operators $b_0$ to the right and
$N_c$ creation operators $b^\dagger_0$ to the left. These contributions are represented diagrammatically by $N_c$ in- and out-going
dashed lines. In contrast,  contractions $\{\Psi_I(x), \bar \Psi_I(y)\}$ at the inner lines contain contributions
from all modes in Eq.~(\ref{def:quark-field-operator-mean-field}), and are represented by solid lines.

\begin{figure}[t]
\begin{center}
\includegraphics[width=12cm,clip=true]{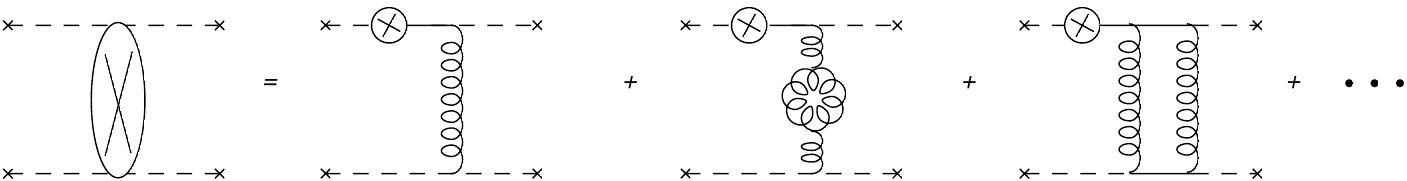}
\end{center}
\caption{Diagrammatic representation of a two-body operator.}
\label{fig:2body_operator_expansion}
\end{figure}

The diagrammatic expansion of Eq.~(\ref{def:QCD-Green-function}) is shown in Fig. \ref{fig:QCD_operator_me_expansion_a} for
a single operator ${\mathcal O}(x) = \bar \Psi(x) \,\Gamma\,\Psi(x)$ with some Dirac flavor matrix $\Gamma$. Multiple gluon
exchanges connect more and more of the dashed lines. As indicated in Fig. \ref{fig:QCD_operator_me_expansion_b}, the various contributions
may be classified by the number of connected dashed lines. Such building blocks involving $r$ connected dashed lines are called $r$-body
operators. In Fig. \ref{fig:2body_operator_expansion} a two-body operator is analyzed in more detail.

The matrix element (\ref{def:QCD-Green-function}) may be expressed in terms of
effective $r$-body operators
\begin{eqnarray}
\bra{\mathcal{B}^\prime}\,\mathcal T\, \mathcal{O}_1(x_1)\ldots \mathcal{O}_n(x_n)\,
\ket{\mathcal{B}} = \sum_l \sum_{r=1}^{N_c} c_r^l (x_1, \ldots , x_n)\, \bra{\mathcal{B}^\prime_0}
\mathcal{O}^{(r)}_l \ket{\mathcal B_0} \,,
\label{def:QCD-operator-Nc-expansion}
\end{eqnarray}
where the first summation extends over all $r$-body operators ${\mathcal O}^{(r)}_l$.
Since a baryon consists of $N_c$ quarks this sum terminates at $r=N_c$. Given $r$, the index $l$ runs over all possible
spin and flavor combinations. The $N_c$-scaling of the unknown functions $c^l_r$,
which parameterize the complicated structure of quark-gluon diagrams, is given by
\begin{eqnarray}
c_r^l \sim  \frac{1}{N_c^{r-1+L}}\,,
\label{def:Nc-scaling}
\end{eqnarray}
where $L$ is the minimal number of quark loops in the diagrams contributing to the matrix elements.
The factor $N_c^{1-r}$, which accompanies each $r$-body operator in the expansion in Eq.~(\ref{def:QCD-operator-Nc-expansion})
is due to the fact, that one needs at least $r-1$ gluon exchanges on the quark-gluon level to generate an $r$-body operator.
Contributions of an $r$-body operator, when evaluated on the
different quark lines, can partially cancel each other. Thus, Eq.~(\ref{def:Nc-scaling}) is an upper bound for the scaling behavior of the
coefficients.

The representation (\ref{def:QCD-operator-Nc-expansion}) does not yet provide the desired $1/N_c$ expansion. This follows from the
scaling behavior of the matrix elements
\begin{eqnarray}
\bra{\mathcal{B}^\prime_0}
\mathcal{O}^{(r)}_l \ket{\mathcal B_0}  \sim  N_c^r \,,
\label{def:scaling-matrix-elements}
\end{eqnarray}
which reflects the combinatorial fact that there are $N_c^r$ possibilities to single out  $r$ lines to form the
$r$-body operator. Again, Eq.~(\ref{def:scaling-matrix-elements}) constitutes an upper bound for the $N_c$ scaling of an $r$-body operator.
Depending on the specifics of the operator, a smaller scaling power may arise.
Thus, it is important to analyze in detail matrix elements of the $r$-body operators with the purpose of reorganizing
the identity  (\ref{def:QCD-operator-Nc-expansion}).

We proceed with the analysis of the $r$-body operators following the work of Dashen, Jenkins and Manohar \cite{Dashen:Jenkins:Manohar:1994}.
A complete set of color-neutral one-body operators is easily constructed with
\begin{eqnarray}
&& \quarknumberoperator = q^\dagger ( \one  \otimes \one  \otimes \one )\,q \,, \qquad  \qquad \;\;
J_i = q^\dagger \Big(\frac{\sigma_i }{2} \otimes \one \otimes \one \Big)\, q \,,
\nonumber\\
&& T^a = q^\dagger \Big(\one \otimes \frac{\lambda_a}{2} \otimes \one \Big)\, q\, ,\qquad \quad \;\;
G^a_i = q^\dagger \Big( \frac{\sigma_i}{2} \otimes \frac{\lambda_a}{2} \otimes \one \Big)\, q\,,
\label{def:one-body-operators}
\end{eqnarray}
where we identify the
operator $q_{\alpha} = b_{0, \alpha}$ with the lowest quark-mode operator with the spin polarization $\alpha =1,2$ as introduced
in Eq.~(\ref{def:quark-field-operator-mean-field}). The one-body operators are normal ordered as required in Eq.~(\ref{def:QCD-operator-Nc-expansion}).
The merit of the one-body operators (\ref{def:one-body-operators}) is their transparent action on the baryon states
\begin{eqnarray}
\bra{\mathcal{B}^\prime_0}\, J_i \,\ket{\mathcal B_0} \sim N_c^0 \,, \qquad
\bra{\mathcal{B}^\prime_0}\, T^a \,\ket{\mathcal B_0} \sim N_c\,, \qquad
\bra{\mathcal{B}^\prime_0}\, G^a_{i}\,\ket{\mathcal B_0} \sim N_c \,.
\label{def:scaling-one-body}
\end{eqnarray}
While the matrix elements of the spin operator scale with $N_c^0$, the matrix elements of the flavor and spin flavor operators may scale with
$N_c$ \cite{Luty:Russel:1994,Dashen:Jenkins:Manohar:1994}.

We turn to the $r$-body operators with $r \geq 2$. Given a product of $k$ one-body operators it decomposes into a sum of normal ordered
$r$-body operators with $r\leq k$ where the expansion coefficients do not depend on $N_c$. Such a decomposition is achieved by successive
internal contractions as implied by the application of the anti-commutator relation
\begin{eqnarray}
\{q_{\alpha, a,A} ,\,q^\dagger_{\beta, b,B}\} = \delta_{\alpha \beta}\,\delta_{ab}\,\delta_{AB}\,,
\label{def:anti-commutation}
\end{eqnarray}
with the spin ($\alpha, \beta$), flavor ($a, b$) and color ($A,B$) indices.  Thus, it appears at first that the coefficients in front of
a product of $k$ one-body operators may scale with $N_c^0$ rather than $N_c^{1-k}$ if the internal contractions give rise to a
contribution of a zero-body operator. However, this does not happen if the normal ordered expression (\ref{def:QCD-operator-Nc-expansion})
is expressed in terms of products of the one-body operators (\ref{def:one-body-operators}). This follows since the
interaction $V_I(t)$ in (\ref{def:QCD-Green-function}) has the particular form
\begin{eqnarray}
V_I(t) = Q^\dagger(t) \,q  + q^\dagger \,Q(t) +
q^\dagger\,G_1(t)\,q +  G_0(t) \,,
\label{def:decompose-V}
\end{eqnarray}
where the operators $G_0(t), G_1(t)$ and $Q(t)$ can involve only gluonic and ghost degrees of freedom and the quark-mode operators of
(\ref{def:quark-field-operator-mean-field}) with the exception of $b_0=q$. If we assume that all operators $\mathcal O_i(x)$
in (\ref{def:QCD-operator-Nc-expansion})
are quark currents, i.e. $\mathcal O_i(x) = \bar \Psi(x)\,\Gamma_i\,\Psi(x)$ with some matrices $\Gamma_i$, it follows that each contribution
to the  r.h.s. of (\ref{def:QCD-operator-Nc-expansion}) in a perturbative expansion is a sum of products of one-body operators at first:
the time ordering on the r.h.s. of  (\ref{def:QCD-Green-function}) cannot alter that structure as is implied by (\ref{def:decompose-V}).
The normal ordered effective $r$-body operators in (\ref{def:QCD-operator-Nc-expansion}) are generated only after
the application of Wick's theorem to the r.h.s. of (\ref{def:QCD-Green-function}). We conclude that if the r.h.s. of
(\ref{def:QCD-operator-Nc-expansion}) is reorganized into a sum of $k$ products of one-body operators, the resulting coefficient
functions scale with $N_c^{1-k-L}$.

Thus, it is useful to expand Eq.~(\ref{def:QCD-operator-Nc-expansion}) into powers of the
one-body operators introduced in Eq.~(\ref{def:one-body-operators}). Matrix elements
of the product of $k$ one-body operators scale with $N_c^{k_{\rm eff}}$,
where $k_{\rm eff}\leq k$ is the number of flavor or spin-flavor operators involved. This follows from Eq.~(\ref{def:scaling-one-body})
by insertion of a complete set of baryon states in between the one-body operators. Since the matrix elements of
the product of $k$ flavor or spin-flavor operators may scale with $N_c^k$, the suppression factor from its associated
coefficients, $N_c^{1-k-L}$, does not provide a truncation and an infinite number of terms had to be considered
at a given order in the $1/N_c$ expansion. However, as proved
in Ref.~\cite{Dashen:Jenkins:Manohar:1994}, such an expansion is tractable owing to the existence of the following two reduction rules:
\begin{itemize}
\item All operator products in which two flavor indices are contracted using $\delta_{ab}$,
$f_{abc}$ or $d_{abc}$ or two spin indices on $G$'s are contracted using $\delta_{ij}$ or $\varepsilon_{ijk}$
can be eliminated.
\item All operator products in which two flavor indices are contracted using symmetric or antisymmetric
combinations of two different $d$ and/or $f$ symbols can be eliminated. The only exception to this rule is the antisymmetric
combination $f_{acg}\,d_{bch}-f_{bcg}\,d_{ach}$.
\end{itemize}
The truncation of the infinite tower of spin-flavor operators follows from a set of operator identities and from the suppression of the spin-operator $J$ in Eq.~(\ref{def:scaling-one-body}).
The only relevant operator identities are those which
relate products of two one-body operators to one-body and zero-body operators.
Identities for all products of $k$ one-body operators with $k>2$ are obtained by recursively applying the two-body identities on all pairs
of one-body operators.

A product of two one-body operators can always be written as a sum of an antisymmetric and a symmetric product using
a commutator or an anticommutator, respectively. The commutator of two one-body operators can always be eliminated
(reduced to a one-body operator) by using the Lie-algebra relations
\begin{eqnarray}
&&\Big[J_i,\, J_j \Big]=i\,\varepsilon_{ijk}\, J_k, \qquad \Big[T^a,\, T^b\Big]=i\,f_{abc}\, T^c, \qquad
\Big[J^i,\, T^a\Big]=0\,,
\nonumber \\
& &\Big[J_i,\, G^{a}_j\Big]=i\,\varepsilon_{ijk}\, G^{a}_k\,,\qquad
\Big[T^a,\, G^{b}_i\Big]=i\,f_{abc}\, G^c_{i}\,,
\nonumber \\
&&\Big[G^{a}_i,\, G^{b}_j\Big]=\frac{i}{4}\,\delta_{ij}\, f_{abc}\, T^c + \frac{i}{6}\, \delta^{ab}\,
\varepsilon_{ijk}\, J_k + \frac{i}{2}\, \varepsilon_{ijk}\, d_{abc}\, G_{k}^c\,,
\label{def:SU(6)-Lie-algebra}
\end{eqnarray}
which may be verified by an explicit computation as a consequence of the canonical anti-commutator relations in Eq.~(\ref{def:anti-commutation}).
It remains to consider the symmetric products of two one-body operators.
An application of the results of Ref.~\cite{Dashen:Jenkins:Manohar:1994} provides the explicit  identities for the symmetric combinations
\begin{eqnarray}
\{T_a, \,T^a\} &=& \frac 16\, (N_c+6)\, \quarknumberoperator + \{J_k,\, J_k\}\,,
\nonumber \\
d_{abc}\, \{T^a, T^b\} &=& -\frac 13\, (N_c+3)\, T^c\,,
\label{result:reduction-identities:1}
\end{eqnarray}
and
\begin{eqnarray}
\{T_a,\, G^{a}_i\} &=& \frac 23\,(N_c+3)\, J_i\,,
\nonumber \\
d_{abc}\, \{T^a, G^{b}_i\} &=& \frac 13\, (N_c+3)\, G^c_i + \frac 16 \{J_i, \,T^c\}\,,
\nonumber \\
f_{abc} \,\{T^a, G^{b}_i\} &=& \varepsilon_{ijk}\, \{J_j, \,G^c_{k}\}\,,
\label{result:reduction-identities:2}
\end{eqnarray}
and
\begin{eqnarray}
\{G^{a}_i, \,G_{ja}\} &=& \frac{1}{8}\,\delta_{ij}\, \Big( (N_c+6)\, \quarknumberoperator
- 2\, \{J_k,\, J_k\} \Big) + \frac 13 \, \{J_i, J_j\}\,,
\nonumber \\
d_{abc}\,\{G^{a}_i, \,G^{b}_j\} &=& \frac 13\,\delta_{ij}\, \Big( \frac 43 \,(N_c+3)\, T^c - \frac 32 \,\{J_k,\, G^c_k\} \Big)
\nonumber \\
&+&\frac 16 \,\Big( \{J_i, \,G^c_j \} + \{J_j,\, G^c_i\} \Big)\,,
\nonumber \\
 \{G^{a}_k,\, G^{b}_k\} &=& \frac{1}{24}\, \delta_{ab}\,\Big( (N_c+6)\, \quarknumberoperator - 2\, \{J_k, J_k\} \Big)
\nonumber \\
&+& \frac 12 \,d_{abc}\,\Big( (N_c+3)\, T^c - 2\, \{J_k, \,G^c_k\} \Big) + \frac 14 \,\{T^a, T^b\}\,,
\nonumber \\
 \varepsilon_{ijk}\,\{G^{a}_j,\, G^{b}_k\}
&=& \frac 12\,  f_{abc}\,\Big(- (N_c+3)\, G^c_{i}+ \frac{1}{6} \,\{J_i,\, T^c\} \Big)
\nonumber \\
&+& \frac 12\, \Big(f_{acg}\,d_{bch}-f_{bcg}\,d_{ach}\Big)\, \{T^g,\, G_{i}^h\} \,,
\label{result:reduction-identities}
\end{eqnarray}
which hold in matrix elements of the baryon ground-state tower. The identities (\ref{result:reduction-identities}) together with the $SU(3)$ relations for the $f$- and $d$-symbols
\begin{eqnarray}
d_{agc}\,d_{bhc} + d_{ahc}\,d_{bgc} &=& \frac 13 \,\big( \delta_{ab}\,\delta_{gh} + \delta_{ag}\,\delta_{bh}
+ \delta_{ah}\,\delta_{bg} \big) - d_{abc} \,d_{ghc} \,,
\nonumber \\
d_{agc}\,d_{bhc} - d_{ahc}\,d_{bgc} &=& f_{abc} \,f_{ghc}
- \frac 23 \,\big(\delta_{ag}\,\delta_{bh} - \delta_{ah}\,\delta_{bg}\big)\,,
\nonumber \\
f_{agc} \,f_{bhc} + f_{ahc} \,f_{bgc} &=& \delta_{ab} \,\delta_{gh} - \delta_{ag}\,\delta_{bh}
- \delta_{ah}\,\delta_{bg} + 3\, d_{abc}\,d_{ghc}\,,
\nonumber \\
f_{agc} \,f_{bhc} - f_{ahc} \,f_{bgc} &=& f_{abc}\, f_{ghc}\,,
\nonumber \\
f_{agc} \,d_{bhc} + f_{ahc} \,d_{bgc} &=& f_{abc}\, d_{ghc}\,,
\label{result:f-d-symbols-identities}
\end{eqnarray}
prove the two reduction rules formulated first in Ref.~\cite{Dashen:Jenkins:Manohar:1994}.

\section{Analysis of matrix elements of the product of two axial-vector currents}

We turn to the analysis of the two-body counterterms introduced in Eq.~(\ref{def-Q2-terms}). A systematic large-$N_c$ scaling analysis
is performed by a study of their contribution to baryon matrix elements of the product of two axial-vector currents,
\begin{eqnarray}
&& A_\mu^{(a)}(x) = \bar \Psi (x)\,\gamma_\mu \,\gamma_5\,\frac{\lambda_a}{2}\,\Psi(x) \,,
\label{def-amu}
\end{eqnarray}
where we recall the definition of the axial-vector current of QCD in terms of the Heisenberg quark-field operators $\Psi(x)$.
Given the chiral Lagrangian, it is well defined how to derive the contribution of a given term to such matrix elements. The classical
matrix of source functions, $a_\mu(x)$, enters the chiral Lagrangian via the building block
\begin{eqnarray}
U_\mu =  \frac{i}{2\,f}\,\partial_\mu \,\Phi - i\,a_\mu + \cdots \,.
\label{result:Umu}
\end{eqnarray}
Compute all diagrams  as implied by the chiral Lagrangian that involve an in- and out-going baryon state.
The second functional derivative with respect to $a_\mu (x)$  of a given term defines its contribution to baryon matrix elements of
the time-ordered product of two axial-vector currents
\begin{eqnarray}
C^{(ab)}_{\mu \nu} (q) = i\,\int d^4 x \,e^{-i\,q\cdot x}  \,{\mathcal T}\,A^{(a)}_\mu (x)\,A^{(b)}_\nu(0) \,.
\label{def:axial-axial}
\end{eqnarray}
The baryon octet and decuplet states
\begin{eqnarray}
\ket{p, \,\chi,\, a }\,, \qquad \qquad \ket{p, \,\chi, \,ijk } \,,
\label{def-states}
\end{eqnarray}
are specified by the momentum $p$ and the flavor indices $a=1,\cdots ,8$ and $i,j,k=1,2,3$. The spin-polarization
label is $\chi = 1,2$ for the octet and $\chi =1,\cdots ,4$ for the decuplet states.

Our strategy is to first evaluate  baryon matrix elements of the correlation function (\ref{def:axial-axial}) with the help of the chiral
Lagrangian. Then we work out the $1/N_c$ expansion of the matrix elements of Eq.~(\ref{def:axial-axial}) in application of
Eq.~(\ref{def:QCD-operator-Nc-expansion}). Finally, we perform a matching of the two results.
A correlation of the parameters of the chiral Lagrangian arises.

Since each parameter in the chiral Lagrangian can be dialed freely
without violating any chiral Ward identity, we may focus on the contributions of the two-body counterterms  (\ref{def-Q2-terms}).
We evaluate the leading order contributions to the matrix elements of Eq.~(\ref{def:axial-axial}) for the baryon-octet and baryon-decuplet states
\begin{eqnarray}
&&\langle \bar p, \bar \chi, d |  \,C^{(ab)}_{\mu \nu} (\bar p - p) \,| p, \chi, c  \rangle  = \bar u(\bar p, \bar \chi)\,\Bigg\{
\frac{1}{2}\,g_{\mu \nu} \,
\Big( \Big[ g_{0}^{(S)}+\frac{2}{3}\,g_{D}^{(S)}\Big]\,\delta_{cd}\,\delta_{ab}
\nonumber\\
&& \qquad \qquad + \, \frac{1}{2}\,g_{1}^{(S)}\,\Big[ \delta_{ad}\, \delta_{bc}+\delta_{ac}\,\delta_{bd} \Big] + g_{D}^{(S)} \,d_{abe}\,d_{cde} + g_{F}^{(S)} \,i\,d_{abe}\,f_{cde}
\Big)
\nonumber\\
&& \qquad +\, \frac{i}{2}\,\sigma_{\mu \nu} \,\Big(
 \frac{1}{2}\,g_{1}^{(T)}\,\Big[\delta_{ad}\, \delta_{bc}-\delta_{ac}\,\delta_{bd}\Big]  + g_{D}^{(T)} \,i\,f_{abe}\,d_{cde} - g_{F}^{(T)} \,f_{abe}\,f_{cde} \Big)
\nonumber\\
&& \qquad +\, \frac{1}{4} \left(\gamma_\mu\, (p+\bar p)_\nu + (p+\bar p)_\mu\,\gamma_\nu \right)\Big( \Big[
g_{0}^{(V)}+\frac{2}{3}\,g_{D}^{(V)}\Big]\,\delta_{cd}\,\delta_{ab}
+  \frac{1}{2}\,g_{1}^{(V)}\,\Big[ \delta_{ad}\, \delta_{bc}+\delta_{ac}\,\delta_{bd}\Big]
\nonumber\\
&& \qquad \qquad + g_{D}^{(V)} \,d_{abe}\,d_{cde} + g_{F}^{(V)} \,i\,d_{abe}\,f_{cde}
 \Big)  \Bigg\} \,u(p, \chi)\,,
\label{result:matrix-octet-octet}
\end{eqnarray}
and
\begin{eqnarray}
&&\langle \bar p, \bar \chi, nop |  \,C^{(ab)}_{\mu \nu} (\bar p - p) \,| p, \chi, c  \rangle  = \bar u_\mu (\bar p, \bar \chi)\,\Bigg\{
\frac{1}{16\sqrt 2}\gamma_\nu \gamma_5\,\Big( f_1^{(A)}\, 2d_{abe}\, \Lambda^{nop}_{ce} + f_2^{(A)}\, 2if_{abe}\, \Lambda^{nop}_{ce} \nonumber \\
&&\quad +\,  f_3^{(A)}\, \left[ \Lambda^{nop}_{ae}\, (d_{bce} + if_{bce})  + \Lambda^{nop}_{be}\, (d_{ace} + if_{ace}) \right]\nonumber \\
&&\quad +\,  f_4^{(A)}\, \left[ \Lambda^{nop}_{ae}\, (d_{bce} + if_{bce})  - \Lambda^{nop}_{be}\, (d_{ace} + if_{ace}) \right] \Big)
\Bigg\} \,u(p, \chi) + (\mu\leftrightarrow \nu, a\leftrightarrow b) \,,
\label{result:matrix-decuplet-octet}
\end{eqnarray}
and
\begin{eqnarray}
&&\langle \bar p, \bar \chi, nop |  \,C^{(ab)}_{\mu \nu} (\bar p - p) \,| p, \chi, klm  \rangle  = - \bar u_\alpha (\bar p, \bar \chi)\,\Bigg\{
\frac 14 \, g^{\alpha \beta}\, g_{\mu\nu}\, \Big( \Big[ h_1^{(S)} + \frac 23 h_3^{(S)} + h_5^{(S)} \Big]\, 2\delta^{ab}\, \delta_{klm}^{nop} \nonumber \\
&& \qquad \qquad +\, \Big[2\, h_3^{(S)} + 3\, h_5^{(S)} \Big] \,d_{abe}\, \delta^{nop}_{xyz}\, \Lambda^{e, xyz}_{klm} - \frac 32\,  h_5^{(S)} \, \delta^{nop}_{rst}\,  \big( \Lambda^{a, rst}_{xyz}\, \Lambda^{b, xyz}_{klm} + \Lambda^{b, rst}_{xyz}\, \Lambda^{a, xyz}_{klm} \big) \Big) \nonumber \\
&& \quad +\, \frac 18\, (g^\alpha_\mu\, g^\beta_\nu + g^\alpha_\nu\, g^\beta_\mu )\, \Big(\Big[ h_2^{(S)} + \frac 23 h_4^{(S)} + h_6^{(S)} \Big]\, 2\delta^{ab}\, \delta^{klm}_{nop}  \nonumber \\
&& \qquad \qquad +\, \Big[2\, h_4^{(S)} + 3\, h_6^{(S)} \Big] \,d_{abe}\, \delta^{nop}_{xyz}\, \Lambda^{e, xyz}_{klm}\,  - \frac 32\, h_6^{(S)} \delta^{nop}_{rst}\,  \big( \Lambda^{a, rst}_{xyz}\, \Lambda^{b, xyz}_{klm} + \Lambda^{b, rst}_{xyz}\, \Lambda^{a, xyz}_{klm} \big) \Big) \nonumber \\
&& \quad +\, \frac{1}{16}\, g^{\alpha \beta}\, \left(\gamma_\mu (p+\bar p)_\nu + (p+\bar p)_\mu\,\gamma_\nu\right) \, \Big( \Big[h_1^{(V)} + \frac 23\, h_2^{(V)} + h_3^{(V)} \Big]\, 2\delta^{ab}\, \delta^{klm}_{nop} \nonumber\\
&& \qquad \qquad +\,  \Big[2\, h_2^{(V)} + 3\, h_3^{(V)} \Big]\, d_{abe}\, \delta^{nop}_{xyz}\, \Lambda^{e, xyz}_{klm} -\frac 32\,  h_3^{(V)} \, \delta^{nop}_{rst}\,  \big( \Lambda^{a, rst}_{xyz}\, \Lambda^{b, xyz}_{klm} + \Lambda^{b, rst}_{xyz}\, \Lambda^{a, xyz}_{klm} \big) \Big) \nonumber \\
&& \quad +\, \frac{i}{2}\,\sigma_{\mu \nu} \, h_1^{(T)} \, i\, f_{abe}\, \delta^{nop}_{xyz}\,\Lambda^{e, xyz}_{klm} \Bigg\} \,u_\beta(p, \chi) \, ,
\label{result:matrix-decuplet-decuplet}
\end{eqnarray}
with the flavor transition tensors\footnote{
In (\ref{result:matrix-octet-octet}) one is directly led to the given flavor structures by taking traces of different combinations of the Gell-Mann matrices. Terms in (\ref{result:matrix-decuplet-octet}) and  (\ref{result:matrix-decuplet-decuplet}) are obtained by writing out the dot-notation as in (\ref{def:dot-notation}) and performing some trivial algebra. The only non-trivial exception, which is worth to be mentioned, is posed by the flavor contraction in the $f_3^{(A)}$- and $f_4^{(A)}$-terms in the chiral Lagrangian. These contractions were rewritten in (\ref{result:matrix-decuplet-octet}) with the help of the identity
\begin{eqnarray*}
\delta^{nop}_{ijk}\, \varepsilon_{jqr}\, \big( \lambda^{(a)}_{iq} (\lambda^{(b)} \lambda^{(c)})_{kr} \pm \lambda^{(b)}_{iq} (\lambda^{(a)} \lambda^{(c)})_{kr} \big)
= \Lambda^{a,nop}_e\, (d^{bce}+if^{bce}) \pm \Lambda^{b,nop}_e\, (d^{ace}+if^{ace}).
\end{eqnarray*}
}
\begin{eqnarray}
&& \Lambda_{ab}^{klm} = \Big[\varepsilon_{ijk}\, \lambda^{(a)}_{li}\,
\lambda^{(b)}_{mj} \,\Big]_{\mathrm{sym}(klm)}\,,\qquad \quad
\delta^{\,klm}_{\,nop} \;\;= \Big[\delta_{kn}\,\delta_{lo}\,\delta_{mp} \,\Big]_{\mathrm{sym}(nop)}\,,
\nonumber\\
&&  \Lambda^{ab}_{klm} = \Big[\varepsilon_{ijk}\, \lambda^{(a)}_{il}\,
\lambda^{(b)}_{jm} \,\Big]_{\mathrm{sym}(klm)}\,,\qquad \quad
\Lambda^{a,klm}_{nop} = \Big[\lambda^{(a)}_{kn} \delta_{lo} \,\delta_{mp}\Big]_{\mathrm{sym}(nop)}\,.
\label{def:flavor-transition-matrices}
\end{eqnarray}
In Eqs.~(\ref{result:matrix-octet-octet} - \ref{result:matrix-decuplet-decuplet}) the summation over the indices occurring twice in a single term is understood with $e=1, \ldots 8$ and $r,s,t,x,y,z=1,2,3$.
Furthermore, the symbol '$\mathrm{sym}(nop)$' in Eq.~(\ref{def:flavor-transition-matrices}) asks for a symmetrization of the three
indices $nop$, i.e. take the six permutations and divide out a
factor $6$. The Dirac spinors $u(p, \chi)$ and $u_\alpha (p, \chi)$ are normalized with
$ \bar u(p, \bar \chi)\,u(p, \chi) = \delta_{\bar \chi \chi}$ and
$ \bar u_\alpha(p, \bar \chi)\,u^\alpha (p, \chi) = - \delta_{\bar \chi \chi}$. For the latter it holds
$\gamma^\alpha u_\alpha(p,\chi) = p^\alpha u_\alpha (p, \chi) = 0$.

We turn to the $1/N_c$ expansion of the baryon matrix elements of Eq.~(\ref{def:axial-axial}). At leading order in the $1/N_c$ the baryon-octet
and decuplet states  are degenerate. Thus, it is convenient to suppress the reference to the particular state and use in the following
$| p, \chi  \rangle $ as  a synonym for either a baryon-octet or baryon-decuplet state.
The baryon matrix element of two axial-vector currents is expanded in powers of the
one-body operators (\ref{def:one-body-operators}) in application of the operator reduction rules
(\ref{def:QCD-operator-Nc-expansion}). Owing to the identity (\ref{def:QCD-Green-function}), the
physical baryon state $| p, \chi  \rangle $ corresponds in such an expansion to the bare state $| \chi )$ in the sense
of (\ref{def:baryon-ground-state}, \ref{def:baryon-state-time-evolution}). We find
\begin{eqnarray}
 \langle \bar p, \bar \chi |  \,C^{(ab)}_{ij}\,| p, \chi  \rangle & =&
 - \delta_{ij}\,( \bar \chi | \, g_1\,\big( {\textstyle{1\over 3}}\,\delta_{ab}\,\quarknumberoperator+
d_{abc}\,T^c \big) + {\textstyle{1\over 2}}\,g_2\,\{T^a,\,T^b\}\, | \chi )
\nonumber\\
& +& \frac{(\bar p+p)_i \,(\bar p+p)_j}{4M} \,
( \bar \chi | \,g_3\,\big( {\textstyle{1\over 3}}\,\delta_{ab}\,\quarknumberoperator+
d_{abc}\,T^c \big) + {\textstyle{1\over 2}}\,g_4\,\{T^a,\,T^b \}\, | \chi )
\nonumber\\
&+& \epsilon_{ijk}\, f_{abc}\,( \bar \chi | \,g_5\,G^c_k\, | \chi )
+ ( \bar \chi |\,  {\textstyle{1\over 2}}\,g_6\,\{G^{a}_i,\,G^{b}_j\}\, | \chi )
\nonumber\\
& + &( \bar \chi |\, {\textstyle{1\over 2}}\,g_7\,\{G^{a}_j,\,G^{b}_i \}\, | \chi ) + \cdots \,,
\label{QCD-identity-AA}
\end{eqnarray}
where we focus on the space components of the correlation function.
In the course of the construction of the various structures, parity and time-reversal invariance are taken into account.
In addition, we consider terms only that arise in the small-momentum
expansion of Eqs.~(\ref{result:matrix-octet-octet}, \ref{result:matrix-decuplet-octet}, \ref{result:matrix-decuplet-decuplet}) and that are
required for the desired matching. The mass parameter $M$ in Eq.~(\ref{QCD-identity-AA}) is
identified with the common mass the baryon-octet and baryon-decuplet states take at large-$N_c$.


The parameters $g_{1,3,5} \sim N_c^0 $, being related to one-body operators, approach finite values in the large-$N_c$ limit.
In contrast, the parameters $g_{2,4,6,7} \sim N_c^{-1} $ vanish in this limit. According to  Eq.~(\ref{def:scaling-one-body}), an additional $N_c$ dependence results
from the evaluation of the matrix elements in Eq.~(\ref{QCD-identity-AA}).
Thus, the representation in Eq.~(\ref{QCD-identity-AA}) is accurate to order $N_c^{0}$.

We note that according to the reduction rule the operator
\begin{eqnarray}
\epsilon_{ijk}\,\big(f_{acg}\,d_{bch}-f_{bcg}\,d_{ach}\big)\, \{T^g,\, G_{k}^h\}
\label{def:operator-ambiguity}
\end{eqnarray}
should be considered in Eq.~(\ref{QCD-identity-AA}) in addition. However, owing to the last identity in Eq.~(\ref{result:reduction-identities}) and
\begin{eqnarray}
\{G^{a}_i, G^{b}_j\} -  \{G^{a}_j, G^{b}_i\} &=& \varepsilon_{ijk}\, \varepsilon_{klm}\, \{G^{a}_l,\, G^{b}_m\} \,,
\label{def:GG-antisymmetric-combination}
\end{eqnarray}
it follows that the two operators $\{G^{a}_i, \,G^{b}_j\} $ and $ \{G^{a}_j, \,G^{b}_i\} $ in Eq.~(\ref{QCD-identity-AA}) may be
linearly combined to the structure (\ref{def:operator-ambiguity}) modulo operators that are already considered or that are subleading.
Thus, it is allowed to consider the combination (\ref{def:GG-antisymmetric-combination}) instead of Eq.~(\ref{def:operator-ambiguity}).

The particular combination of operators associated with $g_1$ and $g_3$ follows since the one-body operator contribution
results from quark-gluon diagrams
where the flavor matrices $\lambda_a$ and $\lambda_b$ sit on a single quark line leading to the unique flavor structure
$\lambda_a\,\lambda_b = \frac{2}{3}\,\delta_{ab}+ (d_{abc}+i\,f_{abc})\,\lambda_c$. This
implies that the symmetric part of the one-body operator contributes in the combination
${\textstyle{1\over 3}}\,\delta^{ab}\,\quarknumberoperator+
d^{abc}\,T^c $ always. Multiple gluon exchanges cannot modify this structure.

At $N_c=3$ the baryon octet and decuplet states may be labeled by
$\roundket{a,{\bar \chi}}$  and $\roundket{ijk, \chi}$ with $a=1,\cdots ,8$ and $i,j,k=1,2,3$. The spin-polarization
label is $\chi = 1,2$ for the octet and $\chi =1,\cdots ,4$ for the decuplet states. These states are specific realizations of the generic
expression for $\ket{\mathcal B_0}$ in Eq.~(\ref{def:baryon-ground-state}).
We  extend the results of Ref.~\cite{Lutz:Kolomeitsev:2002} and present
\begin{eqnarray}
J_i \,\roundket{a,\chi}&=&\frac{1}{2}\, \sigma^{(i)}_{{\bar \chi} \chi}\, \roundket{a,{\bar \chi}}\,,
\nonumber \\
T^a\, \roundket{b,\chi}&=& i\,f_{bca}\, \roundket{c,\chi}\,,
\nonumber \\
G^{a}_i\, \roundket{b,\chi}&=&  \sigma^{(i)}_{{\bar \chi} \chi}\, \Big(\frac12\,d_{bca} + \frac{i}{3}\, f_{bca}\Big)\,
\roundket{c,{\bar \chi}} + \frac{1}{2\sqrt{2}}\, S^{(i)}_{{\bar \chi} \chi}\, \Lambda_{ab}^{klm}
\, \roundket{klm,{\bar \chi}}\,,
\nonumber \\
\nonumber \\
J_i \,\roundket{klm, \chi}&=& \frac{3}{2}\,\Big(\vec S \,\sigma_i\, \vec S^\dagger \Big)_{{\bar \chi} \chi}\, \roundket{klm, {\bar \chi}},
\nonumber \\
T^a \,\roundket{klm, \chi}&=& \frac{3}{2}\,\Lambda^{a,nop}_{klm}\, \roundket{nop,\chi},
\nonumber \\
G^a_i \,\roundket{klm, \chi}&=& \frac34 \, \Big(\vec S\,\sigma_i\, \vec S^\dagger \Big)_{{\bar \chi} \chi}\,
\Lambda^{a,nop}_{klm}\, \roundket{nop, {\bar \chi}}  +
\frac{1}{2\sqrt{2}} \, \Big(S^{\dagger}_i \,\Big)_{{\bar \chi} \chi}\, \Lambda^{ab}_{klm}\, \roundket{b,{\bar \chi}}\,,
\label{result:one-body-operators}
\end{eqnarray}
with the Pauli matrices $\sigma_i$ and the transition matrices $S_i$ characterized by
\begin{eqnarray}
&& S^\dagger_i\, S_j= \delta_{ij} - \frac{1}{3}\sigma_i \sigma_j \,, \qquad S_i\,\sigma_j - S_j\,\sigma_i = -i\,\varepsilon_{ijk} \,S_k\,,
\qquad \vec S\,  \vec S^\dagger= \one_{(4\times 4)}\,,
\nonumber\\
&& \vec S^\dagger \,  \vec S =2\, \one_{(2\times 2)}\,, \qquad \vec S \,\vec \sigma = 0 \,,\qquad
\epsilon_{ijk}\,S_i\,S^\dagger_j = i\,\vec S \,\sigma_k\,\vec S^\dagger\,.
\label{def:spin-transition-matrices}
\end{eqnarray}

Evaluation of Eq.~(\ref{QCD-identity-AA}) and an explicit matching to the contributions of the two-body terms to the matrix elements of the correlator $C^{(ab)}_{ij}$ in Eqs.~(\ref{result:matrix-octet-octet}-\ref{result:matrix-decuplet-decuplet}) lead to the central result of
our work
\begin{gather}
g^{(S)}_0 = 2\,g_1 - g_2 - \frac 76 \,g_+, \qquad g^{(S)}_1= - 2\, g_2 + \frac 13 \,g_+\,, \qquad
g^{(S)}_D= 3\,g_2 + \frac 12\, g_+, \qquad
\nonumber \\
g^{(S)}_F= 2\,g_1 - \frac 43 \, g_+\,, \qquad
h^{(S)}_1 = 0, \qquad h^{(S)}_2 = 0, \qquad  h^{(S)}_3 = 3\, g_1 + \frac 92\, g_2 - \frac 94 \,g_+\,,
\nonumber\\
h^{(S)}_4 = 3\,g_+, \quad h^{(S)}_5 = -3\, g_2 + \frac 32\, g_+, \qquad h^{(S)}_6 = -3\, g_+\,,
\nonumber\\
g^{(V)}_0 = 2\,g_3 - g_4, \qquad g^{(V)}_1= - 2\,g_4\,,  \qquad
g^{(V)}_D= 3\,g_4, \qquad g^{(V)}_F= 2\, g_3\,,
\nonumber \\
h^{(V)}_1 = 0, \qquad h^{(V)}_2 = 3\, g_3 + \frac 92\, g_4, \quad h^{(V)}_3 = -3\, g_4\,, \qquad
\nonumber \\
g^{(T)}_1= - g_-, \qquad g^{(T)}_D= - g_5 + g_-, \qquad  g^{(T)}_F = -\frac 23\, g_5  + \frac 56 \,g_- \,,\qquad
h^{(T)}_1 = - \frac 32\, g_5  + \frac 34\,  g_- \,,
\nonumber\\
f^{(A)}_1=0, \qquad f^{(A)}_2 = -4\, g_5 + 2\,g_- \,, \qquad
f^{(A)}_3= 2\, g_+, \qquad f^{(A)}_4= 6\, g_- \,,
\end{gather}
with $g_\pm = (g_6 \pm g_7 )/2$.  Our result implies the existence of  the following 18 sum rules
\begin{gather}
g^{(S)}_F = g^{(S)}_0 - \frac 12\, g^{(S)}_1, \qquad h_1^{(S)}=0, \qquad h_2^{(S)}=0\,,  \qquad
h^{(S)}_3 = \frac 32\, g^{(S)}_0 - \frac 94\, g^{(S)}_1 + \frac 12\, g^{(S)}_D\,,
\nonumber \\
h^{(S)}_4 = 3\, \big(g^{(S)}_D+\frac 32\, g^{(S)}_1 \big)\,, \qquad
h^{(S)}_5 = g^{(S)}_D + 3\, g^{(S)}_1, \qquad h^{(S)}_6 = -3\, \big(g^{(S)}_D+\frac 32\, g^{(S)}_1 \big),
\nonumber\\
g^{(V)}_D = -\frac 32\, g^{(V)}_1=-h^{(V)}_3\,, \qquad g^{(V)}_F = g^{(V)}_0 - \frac 12\, g^{(V)}_1\,, \qquad
h^{(V)}_1 =0, \qquad h^{(V)}_2 = \frac 32\, g^{(V)}_0 - 3\, g^{(V)}_1\,,
\nonumber\\
g^{(T)}_F = -\frac 16\, g^{(T)}_1 + \frac 23\, g^{(T)}_D\,, \qquad  h^{(T)}_1 = \frac 34\, (g^{(T)}_1 + 2\, g^{(T)}_D)\,,
\nonumber \\
f^{(A)}_1 = 0, \qquad f^{(A)}_2 = 2\, (g^{(T)}_1 + 2\, g^{(T)}_D)\,, \qquad
f_3^{(A)} = 3\, g_1^{(S)} + 2\, g_D^{(S)} \,, \qquad
f^{(A)}_4 = - 6\, g^{(T)}_1\,.
\end{gather}

\section{Summary}

We derived the sum rules for the $Q^2$ two-body counterterms of the chiral Lagrangian as implied by the large-$N_c$ operator
analysis. There are all together 25 independent terms in the chiral Lagrangian with baryon octet and decuplet fields that
contribute to the meson-baryon scattering process at chiral order $Q^2$. At leading order in the $1/N_c$ expansion we established
the relevance of 7 parameters only. Consequently, there are 18 sum rules for the $Q^2$ counterterms.

\clearpage

\section*{Appendix}

At $N_c=3$ baryon matrix elements of arbitrary products of one-body operators can be obtained by consecutive
application of (\ref{result:one-body-operators}). We consider matrix elements
of the symmetric two-body operators for the baryon states where the results are
written in a form that facilitates the large-$N_c$ operator analysis of the chiral Lagrangian (see
(\ref{result:matrix-octet-octet}-\ref{result:matrix-decuplet-decuplet})).

For the matrix elements involving baryon-octet states only we find
\begin{eqnarray}
\roundbra{d,{\bar \chi}} \{J_i,\, J_j\} \roundket{c,\chi} &=&  \frac 12\, \delta_{ij}\,\delta_{{\bar \chi} \chi}\, \delta_{cd} \,,
\nonumber \\
\roundbra{d,{\bar \chi}} \{J_i,\, T^a\} \roundket{c,\chi} &=&  \sigma^{(i)}_{{\bar \chi} \chi} \,i\, f_{cda} \,,
\nonumber \\
\roundbra{d,{\bar \chi}} \{J_i,\, G^a_j\} \roundket{c,\chi} &=&
 \delta_{ij}\, \delta_{{\bar \chi} \chi}\, \Big\{\frac 12\,d_{cda} + \frac {i}{3}\,f_{cda} \Big\}\,,
\nonumber \\
\roundbra{d, {\bar \chi}} \{T^a,\, T^{b}\} \roundket{c,\chi} &=&
 \delta_{{\bar \chi} \chi}\, \Big\{\delta_{ab}\,\delta_{dc} - (\delta_{ad}\,\delta_{bc}+\delta_{bd}\,\delta_{ca} ) + 3\,d_{abe}\,d_{ecd}\Big\}\,,
\nonumber \\
\roundbra{d, {\bar \chi}} \{T^a,\, G^b_{i}\} \roundket{c,\chi}
&=&  \sigma^{(i)}_{{\bar \chi} \chi}\; \Big\{
\frac 13\, \delta_{ab}\,\delta_{dc} - \frac 13 \, \Big(\delta_{ad}\,\delta_{bc}+\delta_{bd}\,\delta_{ca} \Big) + d_{abe}\,\Big(d_{ecd} + \frac i2\, f_{ecd} \Big)
\nonumber \\
&& \qquad - \, \frac i2\, \Big(d_{ade}\, f_{bce} + f_{ade}\, d_{bce} \Big) \Big\}\,,
\nonumber \\
\roundbra{d, {\bar \chi}} \{G^{a}_i, \,G^{b}_j\} \roundket{c,\chi} &=&
 \delta_{ij}\,\delta_{{\bar \chi} \chi}\,\Big\{\frac{5}{12} \,\delta_{ab}\,\delta_{dc}
 -\frac{1}{12}\,\Big(\delta_{ac}\,\delta_{bd} +\delta_{ad}\, \delta_{bc} \Big)
\nonumber \\
&& \qquad -\,\Big(\frac{1}{4}\,d_{cde}-\frac{2}{3}\,i\,f_{cde}\Big)\,d_{abe} \Big\}
\nonumber\\
& +&\,i\,\epsilon_{ijk}\,\sigma^{(k)}_{{\bar \chi}\chi}\,\Big\{\frac{1}{4}\, \Big(\delta_{ac}\,\delta_{bd}
-\delta_{ad}\, \delta_{bc} \Big)+\Big(\frac{1}{2}\,d_{dce}+\frac{5}{12}\,i\,f_{cde}\Big)\,i\,f_{abe} \Big\}\,, \quad
\label{Appendix:octet-octet}
\end{eqnarray}
where a summation over the indices $e=1, \cdots , 8$ and $k=1, 2, 3$ is understood.
Contractions of $f$ and $d$ symbols of $SU(3)$ were rewritten by means of the identities (\ref{result:f-d-symbols-identities}).

The transition matrix elements from a baryon-octet to a baryon-decuplet state are
vanishing unless at least one spin-flavor operator $G^a_i$ is involved. It holds:
\begin{eqnarray}
\roundbra{nop, {\bar \chi}} \{J_i,\, G^{a}_j\} \roundket{c,\chi} &=& \frac{1}{4\sqrt2}\,\Big( 3\, i\,\varepsilon_{ijk}\,S_k
+ S_i\, \sigma_j + S_j\,\sigma_i \Big)_{{\bar \chi} \chi}\,
\Lambda^{nop}_{ac}  \,,
\nonumber\\
\roundbra{nop, {\bar \chi}} \{T^a, \,G^{b}_i\} \roundket{c,\chi} &=& \frac{i}{2\,\sqrt 2}\, S^{(i)}_{{\bar \chi} \chi}\,
 \Big\{ f_{abd} \,\Lambda^{nop}_{dc} +  2\,f_{acd}\,\Lambda^{nop}_{bd} \Big\}\,,
\nonumber \\
\roundbra{nop, {\bar \chi}} \{G^{a}_i, \,G^{b}_j\} \roundket{c, \chi}
&=&i\,\varepsilon_{ijk}\,S^{(k)}_{{\bar \chi} \chi}\;
\frac{1}{8\sqrt2}\, \Big\{  - \Big(d_{cdb} + \frac 23 \,i\,f_{cdb}\Big)\, \Lambda_{ad}^{nop}
\nonumber \\
&& \qquad +\, \frac 53 \,\big( i\,f_{abd}\, \Lambda_{dc}^{nop} + i\,f_{acd}\, \Lambda_{bd}^{nop} \big)
- (a \leftrightarrow b)  \Big\}
\nonumber \\
&+& \Big(S_i \,\sigma_j + S_j\,\sigma_i \Big)_{{\bar \chi} \chi}\,
 \frac{1}{4\sqrt2}\, \Big\{ \Lambda^{nop}_{ad} (d_{bcd} + if_{bcd}) +\, (a \leftrightarrow b)\Big\}\,, \qquad
\label{Appendix:octet-decuplet}
\end{eqnarray}
with the summation indices $d= 1, \cdots , 8$ and $k = 1, 2, 3$. The flavor-transition tensors $\Lambda_{ab}^{nop}$ and
$\Lambda^{ab}_{nop} $ and the spin-transition matrices $S_i$ were introduced in Eqs.~(\ref{def:flavor-transition-matrices}) and
(\ref{def:spin-transition-matrices}), respectively. The result (\ref{Appendix:octet-decuplet}) relies on (\ref{result:one-body-operators})
and the flavor relations
\begin{eqnarray}
\Lambda^{nop}_{dc} &=& - \Lambda^{nop}_{cd}, \nonumber \\
if_{acd}\, \Lambda^{nop}_{bd} &=& - \frac 12\, d_{abd}\, \Lambda^{nop}_{cd} + \frac 12\, if_{abd}\, \Lambda^{nop}_{cd}  \nonumber \\
&+& \frac 14\, \Big( \Lambda^{nop}_{ad}\, (d_{bcd}+if_{bcd}) + a\leftrightarrow b \Big) -  \frac 34\, \Big( \Lambda^{nop}_{ad}\, (d_{bcd}+if_{bcd}) - a\leftrightarrow b \Big)\, .
\end{eqnarray}

There remain the matrix elements for the baryon-decuplet states:
\begin{eqnarray}
\roundbra{nop, {\bar \chi}} \{J_i,\, J_j\} \roundket{klm, \chi} &=&
\Big\{\frac 92\, \delta_{ij}\,\delta_{{\bar \chi}\chi} -
3\, \Big(S_i\,S^\dagger_j + S_j\,S^\dagger_i \Big)_{{\bar \chi}\chi} \Big\}\, \delta^{nop}_{klm}\,,
\nonumber\\
\roundbra{nop, {\bar \chi}} \{J_i, \,T^a\} \roundket{klm, \chi} &=& \frac94 \,\Big(\vec S\, \sigma_i\,
\vec S^\dagger \Big)_{{\bar \chi} \chi}\, \delta^{nop}_{rst}\, \Lambda^{a,rst}_{klm} \,,
\nonumber \\
\roundbra{nop, {\bar \chi}} \{J_i, \,G^a_j\} \roundket{klm, \chi} &=&
\Big\{\frac 94 \,\delta_{ij}\,\delta_{{\bar \chi}\chi}
-  \frac 32\, \Big(S_i\,S^\dagger_j + S_j\,S^\dagger_i \Big)_{{\bar \chi}\chi}\Big\} \,\delta^{nop}_{rst}\, \Lambda^{a,rst}_{klm}\,,
\nonumber\\
\roundbra{nop, {\bar \chi}} \{T^a, \,T^b\} \roundket{klm, \chi} &=&  \frac{9}{4}\,\delta_{{\bar \chi} \chi}\,
\delta^{nop}_{rst}\, \Big\{ \Lambda^{a,rst}_{xyz}\, \Lambda^{b,xyz}_{klm} + \Lambda^{b,rst}_{xyz}\, \Lambda^{a,xyz}_{klm} \Big\} \,,
\nonumber \\
\roundbra{nop, {\bar \chi}} \{T^a, G^b_i\} \roundket{klm, \chi} &=& \frac 98 \,\big(\vec S\, \sigma_i\,
\vec S^\dagger\big)_{{\bar \chi} \chi}\,
\delta^{nop}_{rst}\, \Big\{ \Lambda^{a,rst}_{xyz}\, \Lambda^{b,xyz}_{klm} + \Lambda^{b,rst}_{xyz}\, \Lambda^{a,xyz}_{klm} \Big\}\,,
\nonumber \\
\roundbra{nop, {\bar \chi}} \{G^a_i,\, G^b_j\} \roundket{klm, \chi}
&=& \delta_{ij}\,\delta_{{\bar \chi}\chi}\,\Big\{
\frac{9}{16} \,\delta^{nop}_{rst}\,\Lambda^{a,rst}_{xyz}\, \Lambda^{b,xyz}_{klm}
  + (a \leftrightarrow b) \Big\}
\nonumber \\
+ \;\,i\,\varepsilon_{ijk}\,\Big(\vec S \,\sigma_k\, \vec S^\dagger\Big)_{{\bar \chi}\chi}\,
&& \!\!\!\! \!\!\!\!\Big\{  \frac{1}{16}\,
 \Lambda_{ac}^{nop}\,\Lambda^{bc}_{klm} + \frac{3}{16}\, \delta^{nop}_{rst}\, \Lambda^{a,rst}_{xyz}\, \Lambda^{b,xyz}_{klm}    -  (a \leftrightarrow b) \Big\}
\nonumber \\
+ \, \Big(S_i\,S^\dagger_j + S_j\,S^\dagger_i \Big)_{{\bar \chi}\chi}\,
&& \!\!\!\! \!\!\!\!\Big\{ \frac{1}{16}\,
 \Lambda_{ac}^{nop}\,\Lambda^{bc}_{klm} -\,\frac 38 \,\,\delta^{nop}_{rst}\, \Lambda^{a,rst}_{xyz}\, \Lambda^{b,xyz}_{klm}
\, +   (a \leftrightarrow b) \Big\}\,,
\label{Appendix:decuplet-decuplet}
\end{eqnarray}
with the spin summation index $k=1,2,3$ and the flavor summation indices $r,s,t,x,y,z = 1, 2,3$ and $c=1, \ldots 8$.
The flavor-transition tensors $\Lambda^{a,xyz}_{klm}$ and $\delta^{xyz}_{klm}$ were introduced in Eq.~(\ref{def:flavor-transition-matrices}).
The derivation of Eq.~(\ref{Appendix:decuplet-decuplet}) relies on the flavor identities
\begin{eqnarray}
%
\Lambda^{nop}_{ac} \Lambda^{bc}_{klm} - \Lambda^{nop}_{bc} \Lambda^{ac}_{klm} &=& 4\, if_{abc}\, \delta^{nop}_{xyz}\, \Lambda^{c, xyz}_{klm}, \nonumber \\
\delta^{nop}_{rst} \left( \Lambda^{a,rst}_{xyz} \Lambda^{b, xyz}_{nop}  - \Lambda^{a,rst}_{xyz} \Lambda^{b, xyz}_{nop} \right) &=& \frac 23	 \, if_{abc}\, \delta^{nop}_{xyz}\, \Lambda^{c, xyz}_{klm}, \nonumber \\
\Lambda^{nop}_{ac} \Lambda^{bc}_{klm} + \Lambda^{nop}_{bc} \Lambda^{ac}_{klm}
&=& 4\, \delta_{ab}\, \delta^{nop}_{klm} + 6\, d_{abc}\, \delta^{nop}_{xyz}\, \Lambda^{c, xyz}_{klm} \nonumber \\
&-& 3\, \delta^{nop}_{rst} \left( \Lambda^{a,rst}_{xyz} \Lambda^{b, xyz}_{klm}  + \Lambda^{a,rst}_{xyz} \Lambda^{b, xyz}_{klm} \right) \,.
\end{eqnarray}

The results  in Eqs.~(\ref{Appendix:octet-octet}-\ref{Appendix:decuplet-decuplet}) together  with Eq.~(\ref{result:one-body-operators}) may be used to verify the operator identities (\ref{result:reduction-identities}) evaluated for the baryon octet and decuplet states at $N_c=3$.

\end{document}